\documentclass[aps,eqsecnum,twocolumn]{revtex4}
\usepackage{epsfig}
\usepackage{graphicx}
\usepackage{bm}

\newcommand{\eq}[1]{\begin{equation} #1 \end{equation}}

\newcommand{\eqa}[1]{\begin{eqnarray} #1 \end{eqnarray}}

\begin{document}

\title{Nuclear Wave Functions for Spin and Pseudospin Partners}

\author{P.J.~Borycki,$^{1,2}$ J.~Ginocchio,$^{3}$
W.~Nazarewicz,$^{1,4,5}$ and M.~Stoitsov$^{1,5-7}$ }
\affiliation{
$^{1}$Department of Physics, University of Tennessee,
Knoxville, Tennessee 37996\\
$^{2}$Institute of Physics, Warsaw University of Technology, ul. Koszykowa 75, PL~00662~Warsaw, Poland\\
$^{3}$Theoretical Division, Los Alamos National Laboratory, Los Alamos, New Mexico 87545\\
$^{4}$Institute of Theoretical Physics, University of Warsaw \\
ul.~Ho\.{z}a 69, PL 00681 Warsaw, Poland\\
$^{5}$Physics Division, Oak Ridge National Laboratory, P.O.~Box 2008, Oak Ridge, Tennessee 37831, USA\\
$^{6}$Joint Institute for Heavy Ion Research, 
Oak Ridge, Tennessee 37831\\
$^{7}$Institute of Nuclear Research and Nuclear Energy, Bulgarian
Academy of Science, Sofia-1784, Bulgaria\\}

\date{\today}

\begin{abstract}
Using relations between wave functions obtained in the framework of
the relativistic mean field theory, we investigate
the effects of pseudospin and spin symmetry breaking on the single
nucleon wave functions in spherical nuclei. In our analysis, we apply both relativistic
and non-relativistic self-consistent models as well as the harmonic oscillator model.
\end{abstract}

\maketitle


\section{Introduction}
Pseudospin symmetry was suggested in the late sixties
\cite{[Hec69],[Ari69]} based on the small energy difference
between nuclear energy levels with quantum numbers {\it
(n,$\ell$,j=$\ell$+1/2)} and {\it (n-1,$\ell$+2,j=$\ell$+3/2)}.
Pseudospin coupling schemes were observed and used to explain
different aspects of nuclear structure
\cite{[Jon71],[Bak72],[Str72],[Rat73]}. In particular, pseudospin
symmetry was considered in the context of deformation \cite
{[Boh82]} and superdeformation \cite{[Dud87]}, magnetic moment
interpretation \cite{[Tro94],[Stu99],[Stu02]} and identical bands
\cite{[Naz90a],[Ste90]}. However, the origin of this symmetry was
unknown. It was noticed that, in the relativistic mean field (RMF)
calculations, energy differences between pseudospin partners were
small \cite{[Blo95]}, but it was not until 1997 when the origin of
this phenomenon was understood \cite{[Gin97]} in terms of
near-equility of the absolute values of scalar $V_S(\bf r)$ and
vector $V_V(\bf r)$ potentials \cite{[Coh91]}. This observation
stimulated a number of works on this subject
\cite{[Gin98a],[Gin98],[Men98b],[Sug98],[Gin99a],[Gin99],
[Men99],[Sug02],[Gin00a],[Gin01],[Gin01a],[Gin02]}.

Although the origin of pseudospin symmetry has been explained, it
is still an open question why this symmetry is obeyed so well in
non-relativistic calculations. This question is particularly
important for superheavy nuclei and nuclei far from stability
where shell effects are crucial and a subtle change in interaction
may effect even in changing magic numbers (see, e.g., discussion
in Ref.~\cite{[Kru00]}).

Although spin symmetry appears to be broken since spin-orbit
splittings are very large, it is possible that the spin symmetry
breaking may be a dynamic symmetry, i.e.,  the energy levels are
not degenerate but the eignefunctions preserve the symmetry. In
this paper we, therefore, compare effects of both pseudospin and
spin symmetry breaking for non-relativistic as well as
relativistic single-nucleon eigenfunctions for spherical nuclei.

The structure of the paper is the following. In Sec.~\ref{sec2},
we briefly introduce the main pseudospin relations consistent with the
Dirac equation. Numerical evidences for pseudospin dynamic
symmetry in nuclei are analyzed in Sec.~\ref{sec3}, while the
Sec.~\ref{sec4} is devoted to the spin symmetry. Calculations have been performed for a number of spherical doubly-magic nuclei. The results presented in this paper correspond to $^{208}$Pb, as a representative case. Conclusions  are
presented in Sec.~\ref{sec5}.

\section{Pseudospin symmetry and the Dirac Hamiltonian}
\label{sec2}

\subsection{Pseudospin Conditions on the Dirac Hamiltonian}
\label{sec21}

The Dirac Hamiltonian ($\hbar=c=1$)
\begin{equation}
H = \boldsymbol{\alpha} \cdot {\bf p}
    + V_V({\bf r}) + \beta V_S({\bf r}) + \beta\ M.
\label{Hs}
\end{equation}
with  external scalar $V_S(\bf {r})$ and vector $V_V({\bf r})$
potentials, vanishing space components and non-vanishing time
component, is invariant under an SU(2) algebra if the scalar
potential $V_S({\bf r})$ and the vector potential $V_V({\bf r})$
are related up to a constant $C_{ps}$ \cite {[Bel75],[Gin97]}:
\begin {equation}
V_S({\bf r}) + V_V({\bf r}) = C_{ps}. \label{rel}
\end {equation}
The pseudospin generators
\begin{equation}
{\bf {\tilde S}} = \left( \begin{array}{cc}
{\bf {\tilde s}} & 0 \\
0 & {\bf { s}} \end{array}
     \right)
\label{ps}
\end{equation}
form an SU(2) algebra
\begin{equation}
[{\tilde S}_i,{\tilde S}_j] = i \epsilon_{ijk}\ {\tilde S}_k,
\end{equation}
where ${\bf {\tilde s}} = U_p\ {\bf { s}}\ U_p~$, $U_p = \boldsymbol{\sigma}
\cdot \hat{\bf{p}}$ is the helicity transformation \cite {[Blo95]},
${\bf s} =\boldsymbol{\sigma}/2$, and ${\sigma_i}~$ are the
usual Pauli matrices.

The operators ${\tilde S}_i$ commute with the Dirac Hamiltonian
$H_{ps}$ satisfying conditions (\ref{rel}):
\begin{equation}
[{\tilde S}_i, H_{ps}] =0,
\label {comm}
\end{equation}
thus  generating an SU(2) invariant symmetry of $H_{ps}$.

\subsection{Pseudospin Conditions on the Dirac Eigenfunctions}
\label{sec22}

According to the  SU(2) invariant symmetry of $H_{ps}$, each
eigenstate of the Dirac Hamiltonian $H_{ps}$ with the third component of pseudospin ${\tilde \mu} =
 {1\over 2}$ has a partner with ${\tilde \mu} = -
{1\over 2}$ and the same energy, i.e.,
\begin{equation}
H_{ps}\ \Phi_{k{\tilde \mu}}^{ps}({\bf r}) = E_k \Phi_{k{\tilde
\mu}}^{ps}({\bf r}) \label{eigen}
\end{equation}
where  ${\tilde \mu} = \pm {1\over 2}$ is the eigenvalue of
${\tilde S}_z$,
\begin{equation}
{\tilde S}_z\ \Phi_{k{\tilde \mu}}^{ps}({\bf r}) = {\tilde \mu} \
\Phi_{k{\tilde \mu}}^{ps}({\bf r}), \label{Sz}
\end{equation}
and $k$ are the other quantum numbers. The eigenstates in the
doublet are connected by the generators ${\tilde S}_{\pm}~$:
\begin{equation}
{\tilde S}_{\pm }\  \Phi_{k{\tilde \mu}}^{ps}({\bf r}) =
\sqrt{{\left ({1 \over 2} \mp {\tilde \mu} \right )\left({3 \over 2}
\pm {\tilde \mu}\right)}}  \ \Phi_{k{\tilde \mu} \pm 1}^{ps}({\bf r}).
\label{S+}
\end{equation}

The generators (\ref {Sz}), (\ref {S+}) do not mix upper and lower
components of the Dirac eigenfunctions $\Phi_{k{\tilde
\mu}}^{ps}({\bf r})$. Since the spin operates on the lower
components only (see Eq. (\ref{ps})), one of predictions of this symmetry is that
the spatial amplitudes of the lower components of the Dirac
wave functions are identical in shape \cite {[Gin98]}. For
spherical nuclei this means that the lower components of the
pseudospin doublets have the same radial quantum number ${\tilde
n} = n $ \cite {[Gin01a]} and the same spherical harmonic rank
${\tilde \ell} = \ell + 1$. Therefore, it is natural to label the
doublets with their pseudospin quantum numbers $({\tilde
n},{\tilde \ell}, j = {\tilde \ell} \pm {1 \over 2})$. The Dirac
eigenfunctions then have the form
\begin{equation}
\Phi^{ps}_{{\tilde n}, {\tilde \ell},j, m}({\bf r}) = \ \left(
\begin{array}{c}
g_{{\tilde n}{\tilde \ell}j}(r)\ [Y^{({\tilde \ell}_j)} (\theta,
\phi)\ \chi]^{(j)}_m
\\ if_{{\tilde n}{\tilde \ell}}(r) \ [Y^{({\tilde \ell})} (\theta,
\phi)\ \chi]^{(j)}_m
\end{array}
\right).
\end{equation}
where $\chi_{\mu}$ is the spin function, $Y^{({\tilde \ell})}
(\theta, \phi)$ is the spherical harmonic of rank ${\tilde \ell}$,
and ${\tilde \ell}_j = {\tilde \ell} \pm 1$ for $ j = {\tilde
\ell} \pm {1 \over 2}$

As stated above, the radial wave functions of the lower components are equal:
\eq{f_{{\tilde n}{\tilde\ell}j= {\tilde \ell} -
{1\over2}}(r)=f_{{\tilde n}{\tilde \ell}j={\tilde
\ell}+{1\over2}}(r). \label{eq1f:eq}}
On the other hand,
the generators for the upper components depend on the momentum as
well as the spin so they intertwine spin and space. Therefore, in
the pseudospin symmetry limit, the radial wave functions of the upper
components $g_{{\tilde n}{\tilde \ell} j = {\tilde \ell} \pm {1
\over 2}}(r)$ satisfy differential
relations  \cite{[Gin02]}:
\eqa{\lefteqn{D_{{\tilde n}{\tilde\ell} j= {\tilde \ell}
- {1\over 2}}(r)\ g_{{\tilde n}{\tilde
\ell}j={\tilde\ell}-{1\over2}}(r)={}}\nonumber\\ &&{}=D_{\tilde{n}\tilde{\ell} j={\tilde \ell} + {1\over 2}}(r)g_{{\tilde n}{\tilde \ell}
j={\tilde\ell}+{1\over2}}(r),   \label{eq1:eq}}
where 
\eq{\begin{array}{lll} D_{{\tilde n}{\tilde
\ell}j={\tilde\ell}-{1\over2}}(r)&=&\left(\frac{d}{dr}-\frac{{\tilde
\ell}-1}{r} \right), \\ \\ 
D_{{\tilde n}{\tilde
\ell} j =
{\tilde\ell}+{1\over2}}(r)&=&\left(\frac{d}{dr}+\frac{{\tilde
\ell}+2}{r}\right). 
\end{array} \label{eqnLRdf}}

In the  non-relativistic limit, $g_{{\tilde n}{\tilde \ell} j}(r)$
is associated with the single particle wave function while $f_{{\tilde
n}{\tilde \ell} j }(r)$ vanishes. Relativistic mean field
calculations show that indeed $f_{{\tilde n}{\tilde \ell} j }(r)$
is small compared to $D_{{\tilde n}{\tilde \ell} j }g_{{\tilde
n}{\tilde \ell} j }(r)$. The factor is roughly six as we
shall see below.

\subsection{Dirac Conditions on the Pseudospin Doublet States}
\label{sec23}

 Pseudospin symmetry relates upper components in the
doublets to each other and lower components to each other, but
pseudospin symmetry does not relate upper components to lower
components because the pseudospin generators (\ref{ps}) are
diagonal. It is, of course, the Dirac equation, which relates upper to lower
components.

For spherically symmetric potentials, the Dirac equation is reduced
to  coupled first order differential equations in the radial
coordinate only leading to \cite {[Gin97]}
\eqa{\lefteqn{D_{\tilde{n}\tilde{\ell}j}(r)
g_{\tilde{n}\tilde{\ell}j}(r)={}}\nonumber\\&&{}=\left[2M + V_S(r) - V_V(r)-
E\right]f_{\tilde{n}\tilde{\ell}j}(r),\label{Dirac}\\ 
\lefteqn{D_{n{\ell}j}(r)  f_{{ n}{ \ell}j}(r)={}}\nonumber\\&&{}=\left[V_S(r) + V_V(r)+ E\right]g_{{n}{
\ell}j}(r),\label{Dirac2}} 
where $V_{S}( r)$ and $
V_{V}(r)$ are spherical potentials and $E$ is the binding energy.

For heavy nuclei, the vector and scalar potentials are
approximately constant inside the nuclear interior. At the nuclear
surface the potentials fall rapidly to zero and hence outside the
nuclear surface both $f(r)$ and $g(r)$ decrease exponentially.
Also the nucleon mass is very large compared to the binding energy. 
In the nuclear interior $V_S(r) - V_V(r)\approx$const, 
hence 
\eq{D_{{\tilde n}{\tilde
\ell}j}(r)g_{{\tilde n}{\tilde \ell}j}(r) \approx \ \lambda\ f_{{\tilde
n}{\tilde \ell}j}(r), \label{6}}
where
$\lambda=2M+V_S(r)-V_V(r)-E\approx 6$\,fm$^{-1}$ in our calculations.
Notice that in the pseudospin
symmetry limit $f_{{\tilde n}{\tilde \ell}j}(r) = f_{{\tilde
n}{\tilde \ell}}(r)$ and, therefore, Eqs.~(\ref{Dirac}), (\ref{Dirac2}) are
consistent with Eq.~(\ref {eq1:eq}).

\section{Pseudospin Dynamic Symmetries}
\label{sec3}

The relation (\ref{eq1:eq}) is strictly fulfilled only under
condition (\ref{rel}), $V_S+V_V= C_{ps}$, \cite{[Gin02]}.
Therefore, comparing the differences between $D_{{\tilde n}{\tilde
\ell} j={\tilde\ell}-{1\over2}}(r)g_{{\tilde n}{\tilde \ell} j =
{\tilde \ell} - {1 \over 2}}(r)$ and $D_{{\tilde n}{\tilde \ell}
j={\tilde\ell}+{1\over2}}(r)g_{{\tilde n}{\tilde \ell} j = {\tilde
\ell} + {1 \over 2}}(r)$, one can learn about the pseudospin
symmetry breaking effects. The differential relations (\ref
{eq1:eq}) have been checked previously only for the RMF
approximation of a relativistic Lagrangian with zero range
interactions \cite{[Gin02]}. The pseudospin breaking effects have
been studied also by taking the integral form of
Eqs.~(\ref{eq1:eq}) but integral relations depend on the boundary
conditions and hence are less general than the differential
relations \cite{[Gin01]}.

In this work, we investigate the pseudospin breaking effects  for
spherical double-magic nuclei by carrying out three type of
calculations. First, we use the standard harmonic oscillator (HO)
wave functions. Secondly, we  perform non-relativistic
self-consistent Hartree-Fock (HF) calculations with the SLy4 Skryme
force \cite{[Cha98]}. Finally, we  perform relativistic
mean field (RMF) calculations using the Lagrangian \cite{[Ser97]}
with the NL1 parameter set \cite{[Lal97]}.

\subsection{Comparison within the harmonic oscillator model}
\label{sec31}

\begin{figure}
\begin{center}
\includegraphics[width=0.95\columnwidth]{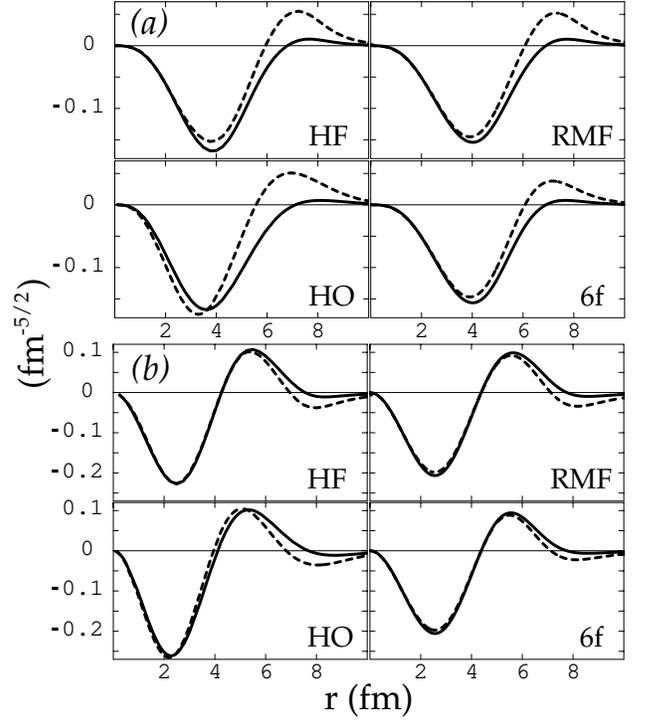}
\end{center}
\caption{Numerical check of identities (\ref{eq1:eq}), (\ref{eq1f:eq}). 
Comparison between $D_{\tilde{n}\tilde{\ell}j=\tilde{\ell}-{1 \over 2}}g_{\tilde{n}\tilde{\ell}j=\tilde{\ell}-{1 \over 2}}$ (dashed line) and $D_{\tilde{n}\tilde{\ell}j=\tilde{\ell}+{1 \over 2}}g_{\tilde{n}\tilde{\ell}j=\tilde{\ell}+{1 \over 2}}$ (solid line)
for $1\tilde{f}$ (a) and $2\tilde{d}$ (b) pseudospin doublets 
in $^{208}$Pb
obtained in different
methods: HF, HO
and RMF. The plot labeled `6f'
shows 6 times scaled lower components of 
the RMF wave function, see Eq. (\ref{6}) and related discussion.}
\label{comparison1:pic}
\end{figure}

For the spherical harmonic oscillator potential (or
spherical Nilsson potential) we take the analitycal form of wave
functions with an oscillator frequency
$\hbar\omega=41/A^{1/3}$ \cite{[Mos57]}. Then, one can express
$D_{{\tilde n}{\tilde\ell}j}(r)g_{{\tilde n}{\tilde \ell} j }(r)$
defined by Eq.~(\ref{eq1:eq}) as:
\eq{ \begin{array}{lll} &&D_{{\tilde n}{\tilde \ell} j= {\tilde
\ell} - {1\over 2}}(r)g_{{\tilde n}{\tilde \ell} j = {\tilde \ell}
- {1 \over 2}}(r)=\\
&&=~\displaystyle{ \chi_{{\tilde n}{\tilde \ell}}(x)\
\sum_{a=0}^{{\tilde n}-1}\frac{(-1)^a(2{\tilde n}+{\tilde
\ell}-3/2-a)x^a}{({\tilde
n}-1-a)!a! \Gamma(a+{\tilde \ell}+3/2)}}, \\ \\
&&D_{{\tilde
n}{\tilde\ell} j= {\tilde \ell} + {1\over 2}}(r)g_{{\tilde
n}{\tilde \ell} j = {\tilde \ell} + {1 \over
2}}(r)=\sqrt{\frac{{\tilde n}+{\tilde \ell}-1/2}{{\tilde n}-1}}\times\\
&&\times~\displaystyle{\chi_{{\tilde n}{\tilde \ell}}(x)\
\sum_{a=0}^{{\tilde n}-1}\frac{(-1)^a(2{\tilde
n}-2-a)x^a}{({\tilde n}-1-a)!a!\Gamma(a+{\tilde \ell}+3/2)}},
\end{array} \label{hogg}}
where the envelope function is
\eq{\chi_{{\tilde n}{\tilde
\ell}}(x)=\sqrt{\frac{2(2\nu)^{\frac{6-{\tilde
\ell}}{2}}(n-1)!}{\Gamma({\tilde n}+{\tilde \ell}-1/2)^3}
}x^{\frac{{\tilde \ell}}{2}} e^{-x/2},}
and
\eq{x=r^22\nu,~~~ \nu=\frac{m\omega}{2\hbar^2}.\label{x:df}}
Expressions (\ref{hogg}) can be presented as products
\eq{D_{{\tilde n}{\tilde \ell}j}(r)g_{{\tilde n},{\tilde \ell}, j
}(r)=\chi_{{\tilde n}{\tilde \ell}}(x)P_{{\tilde
n}{\tilde\ell}j}(x)\label{PLAL1}} 
of the common envelope function
$\chi_{{\tilde n}{\tilde \ell}}(r)$ and a polynomials $P_{{\tilde
n}{\tilde \ell}j}(x)$ with power expansion coefficients
$A_a({\tilde n},{\tilde \ell},j)$: 
\eq{P_{{\tilde n}{\tilde
\ell}j}(x)=\sum_{a=0}^{\tilde{n}-1}(-1)^aA_a({\tilde n},{\tilde \ell},j)\
x^a.\label{PLAL2}} 
As seen from Eq.~(\ref{PLAL2}) these polynomials are of the same order
$({\tilde n} -1)$ independent of $j$, whereas the original harmonic
oscillator eigenfunction with $j = {\tilde \ell} - {1\over 2}$ involves
a polynomial of order ${\tilde n}$ while the harmonic oscillator
eigenfunction with $j = {\tilde \ell} + {1\over 2}$ involves a
polynomial of order  ${\tilde n} - 1$ in $x$.

As an example, in the lower left corners of
Fig.~{\ref{comparison1:pic} (a) and (b) we compare
$D_{{\tilde n}{\tilde \ell}j= {\tilde \ell} +
{1\over2}}(r)g_{{\tilde n}{\tilde \ell} j = {\tilde \ell} + {1
\over 2}}(r)$ (dashed line) with $ D_{{\tilde n}{\tilde \ell}j=
{\tilde \ell} - {1\over2}}(r)g_{{\tilde n}{\tilde \ell} j =
{\tilde \ell} - {1 \over 2}}(r)$ (solid line) using expression
(\ref{hogg}) or its equivalent (\ref{PLAL1}). It is seen that  $ D_{{\tilde
n}{\tilde \ell}j= {\tilde \ell} + {1\over2}}(r)g_{{\tilde
n}{\tilde \ell}j= {\tilde \ell} + {1\over2}}(r)\approx D_{{\tilde
n}{\tilde \ell}j= {\tilde \ell} - {1\over2}}(r)g_{{\tilde
n}{\tilde \ell}j= {\tilde \ell} - {1\over2}}(r)$ in the whole range of $r$ considered. 

\begin{figure}
\begin{center}
\includegraphics[width=0.95\columnwidth]{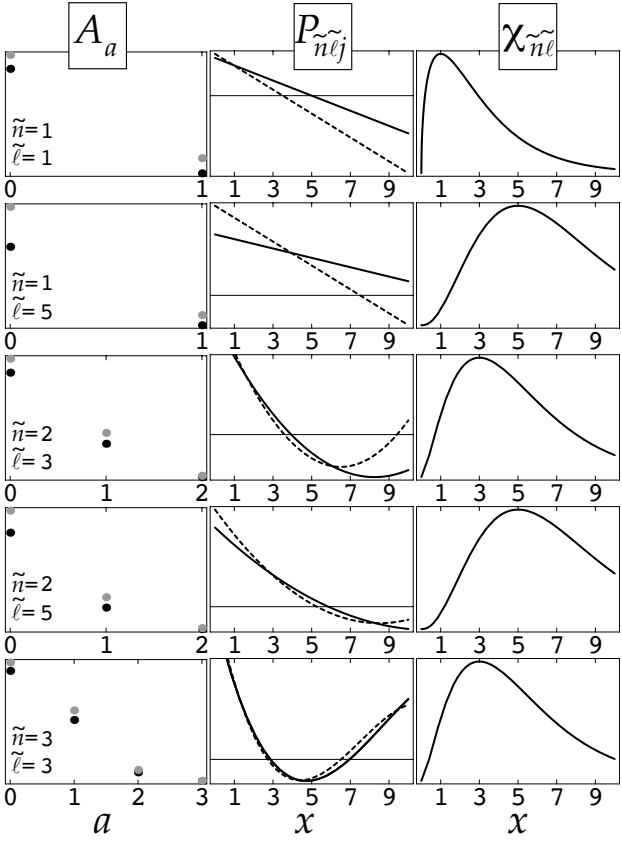}
\end{center}
\caption{Structure of $D_{\tilde{n}\tilde{\ell}j}g_{\tilde{n}\tilde{\ell}j}$ for
 $j$=$\ell$--${1 \over 2}$ (dashed line, grey points), 
 $j$=$\ell$+${1 \over 2}$ (solid line, black points)
obtained in the harmonic oscillator model for different pairs
$\tilde{n},\tilde{\ell}$ in $^{208}$Pb. The first column
shows absolute values of coefficients $A_a$ (Eqs.
(\ref{PLAL1}, \ref{PLAL2})) as a function of $a$. The second column
shows $P_{\tilde{n}\tilde{\ell}}$, third column - envelope functions $\chi_{\tilde{n}\tilde{\ell}}$;
both in terms of dimensionless variable $x$, defined in equation (\ref{x:df}).}
\label{oscillator:pic}
\end{figure}

Systematic calculations of many states and nuclei have shown that
relation (\ref{eq1:eq}) holds better as ${\tilde n}$ increases
or ${\tilde \ell}$ decreases. In order to understand this property
we can use the analytical results from Eq.~(\ref{PLAL1}). For this
purpose plots of $P_{{\tilde n}{\tilde \ell}j}(x),~A_a({\tilde
n},{\tilde \ell},j)$ are presented in Fig.~\ref{oscillator:pic}.
  It is seen that $P_{{\tilde n}{\tilde
\ell}j= {\tilde \ell} - {1\over2}}(x)$ and $P_{{\tilde n}{\tilde
\ell}j= {\tilde \ell} + {1\over2}}(x)$ are overlapping more with
${\tilde n}$ increasing. While the quantum number ${\tilde n}$ is
responsible for the similarities between polynomials, the envelope
functions $\chi_{{\tilde n}{\tilde \ell}}(r)$ strongly depend on
${\tilde \ell}$. The higher ${\tilde \ell}$ is the broader is the
`bell' of $\chi_{{\tilde n}{\tilde \ell}}(r)$. Consequently, when
${\tilde \ell}$ decreases, the differences between $ D_{{\tilde
n}{\tilde \ell}j= {\tilde \ell} + {1\over2}}(r)g_{{\tilde
n}{\tilde \ell}j= {\tilde \ell} + {1\over2}}(r)$ and $ D_{{\tilde
n}{\tilde \ell}j= {\tilde \ell} - {1\over2}}(r)g_{{\tilde
n}{\tilde \ell}j= {\tilde \ell} - {1\over2}}(r)$ are reduced in
the region where $P_{{\tilde n}{\tilde \ell}j= {\tilde \ell} +
{1\over2}}(x)$ and $P_{{\tilde n}{\tilde \ell}j= {\tilde \ell} -
{1\over2}}(x)$ differ.
In general, when $ {\tilde n} \gg {\tilde \ell}$, then $A_a({\tilde
n},{\tilde \ell},j= {\tilde \ell} - {1\over2}) \approx A_a({\tilde
n},{\tilde \ell},j= {\tilde \ell} + {1\over2})$ and hence
$D_{{\tilde n}{\tilde \ell}j= {\tilde \ell} -
{1\over2}}(r)g_{{\tilde n}{\tilde \ell}j= {\tilde \ell} -
{1\over2}}(r) \approx D_{{\tilde n}{\tilde \ell}j= {\tilde \ell} +
{1\over2}}(r)g_{{\tilde n}{\tilde \ell}j= {\tilde \ell} +
{1\over2}}(r)$.

\subsection{Comparison within self--consistent models}

In Figs.~{\ref{comparison1:pic} (a) and (b) we plot
$D_{{\tilde n}{\tilde \ell}j= {\tilde \ell} +
{1\over2}}(r)g_{{\tilde n}{\tilde \ell} j = {\tilde \ell} + {1
\over 2}}(r)$ (dashed line) and $ D_{{\tilde n}{\tilde \ell}j=
{\tilde \ell} - {1\over2}}(r)g_{{\tilde n}{\tilde \ell} j =
{\tilde \ell} - {1 \over 2}}(r)$ (solid line) using the HO model, the non-relativistic 
HF approximation, and the RMF approximation.

Comparing the non-relativistic and relativistic mean field results
we see that the agreement is comparable or slightly better than
the harmonic oscillator results. Therefore, the pseudospin
symmetry relations are not only approximately valid for the
relativistic mean field eigenfunctions but also for the
non-relativistic HF eigenfunctions. Hence we seem to
have pseudospin dynamic symmetry, that is, the energy levels are
not degenerate but the eigenfunctions preserve the pseudospin
symmetry.

In order to confirm that the radial wave functions of the lower
components are approximately equal within a doublet we also plot
them in the case of RMF calculations in the lower right corner of
Figs.~{\ref{comparison1:pic} (a) and (b). We
multiply these wave functions by a factor of 6 in order to be
comparable to the upper components as suggested by Eq.~(\ref {6}).
Indeed the amplitudes of the lower components are approximately
equal \cite {[Gin98]}.

\section{Spin symmetry and the Dirac Hamiltonian}
\label{sec4}

\subsection{Spin Conditions on the Dirac Eigenfunctions}
\label{sec41}

The Dirac Hamiltonian is invariant under an SU(2) algebra if the
scalar potential, $V_S({\bf r})$, and the vector potential
$V_V(\vec{r_i})$, are related \cite {[Bel75],[Gin97],[Pag01]}:
\begin {equation}
V_S({\bf r}) - V_V({\bf r}) = C_{s}, \label{relspin}
\end {equation}
where $C_{s}$ is a constant. Hence spin symmetry can occur for
very relativistic systems like quarks in a meson where both
$V_S({\bf r})$ and $V_V({\bf r})$ are large \cite {[Pag01]}.

The spin generators
\begin{equation}
{\bf {S}} = \left( \begin{array}{cc}
{\bf { s}} & 0 \\
0 & {\bf {\tilde s}} \end{array}
     \right)
\label{s}
\end{equation}
form an SU(2) algebra
\begin{equation}
[{ S}_i,{ S}_j] = i \epsilon_{ijk}\ { S}_k,
\end{equation}
and commute with the Dirac Hamiltonian $H_{s}$ satisfying conditions
(\ref{relspin})
\begin{equation}
[{ S}_i, H_{s}] =0.
\label {comms}
\end{equation}

Thus the operators ${ S}_i$ generate an SU(2) invariant symmetry
of $H_{s}$. Therefore, each eigenstate of the Dirac Hamiltonian
$H_{s}$ has a  partner with the same energy,
\begin{equation}
H_{s}\ \Phi_{k{ \mu}}^{s}({\bf r}) = E_k \Phi_{k{ \mu}}^{s}({\bf
r}), \label{eigens}
\end{equation}
where $k$ are the other quantum numbers and ${ \mu} = \pm
{1\over 2}$ is the eigenvalue of
${ S}_z$,
\begin{equation}
{ S}_z\ \Phi_{k{ \mu}}^{s}({\bf r}) = { \mu} \ \Phi_{k{
\mu}}^{s}({\bf r}). \label{Spz}
\end{equation}
The eigenstates in the spin doublet will be connected by the
generators ${ S}_{\pm}$,
\begin{equation}
{ S}_{\pm }\  \Phi_{k{ \mu}}^{s}({\bf r}) = \sqrt{{\left ({1 \over
2} \mp { \mu} \right )  ({3 \over 2} \pm { \mu} )}}  \ \Phi_{k{
\mu} \pm 1}^{s}({\bf r}). \label{Sp+}
\end{equation}

The generators (\ref {Spz}), (\ref {Sp+}) do not mix upper and
lower components. Since the spin operates on the upper
components only, one of predictions of this symmetry is that the
radial wave functions of the upper components of the Dirac
eigenfunctions are identical. In the spherical symmetry
limit the Dirac eigenfunctions then have the form
\begin{equation}
\Phi^{s}_{{ n}_r, { \ell},j, m}({\vec r}) = \ \left(
\begin{array}{c}
g_{n\ell}(r)\ [Y^{({ \ell})} (\theta, \phi)\ \chi]^{(j)}_m
\\ if_{n\ell j}(r) \ [Y^{({ \ell}_j)} (\theta,
\phi)\ \chi]^{(j)}_m
\end{array}
\right),
\end{equation}
where ${ \ell}_j = {\ell} \pm 1$ for $ j = { \ell} \pm {1 \over 2}$.

The spin generators (\ref {s}) are related to the pseudospin generators by
$\bf{S} = \gamma_5 \bf{\tilde S} \gamma_5$ where $\gamma_5
  = \left( \begin{array}{cc}
{0} & 1 \\
1 & 0 \end{array}
     \right).$ Therefore, the conditions are the same as for the pseudospin
except that now ${\tilde n} \rightarrow n, {\tilde \ell}
\rightarrow \ell, g(r) \rightarrow f(r), f(r) \rightarrow g(r)$:
\eqa{\lefteqn{D_{{ n}{\ell} j= { \ell} - {1\over 2}}(r)\
f_{{ n}{ \ell}j={\ell}-{1\over2}}(r)={}}\nonumber\\&&{}=D_{n\ell j= { \ell} +
{1\over 2}}(r)f_{{ n}{ \ell} j={\ell}+{1\over2}}(r)}
and
\eq{g_{{
n}{ \ell}j= { \ell} - {1\over2}}(r)=g_{{ n}{ \ell}j={
\ell}+{1\over2}}(r). \label{eq1:eqs}}
For finite nuclei $C_s=0$ in Eq.~(\ref{relspin}) because
the potentials go to zero for large $r$. Consequently, equality
$V_S(r) = V_V(r) $ emerges in the spin symmetry limit. Therefore,
we do not expect spin symmetry to be conserved in nuclei since it is
known that $V_{S}(r)$ and $V_{V}(r)$ are both large and of
opposite sign.

\subsection{Spin breaking for different models}
\label{sec42}

\begin{figure}
\begin{center}
\includegraphics[width=0.95\columnwidth]{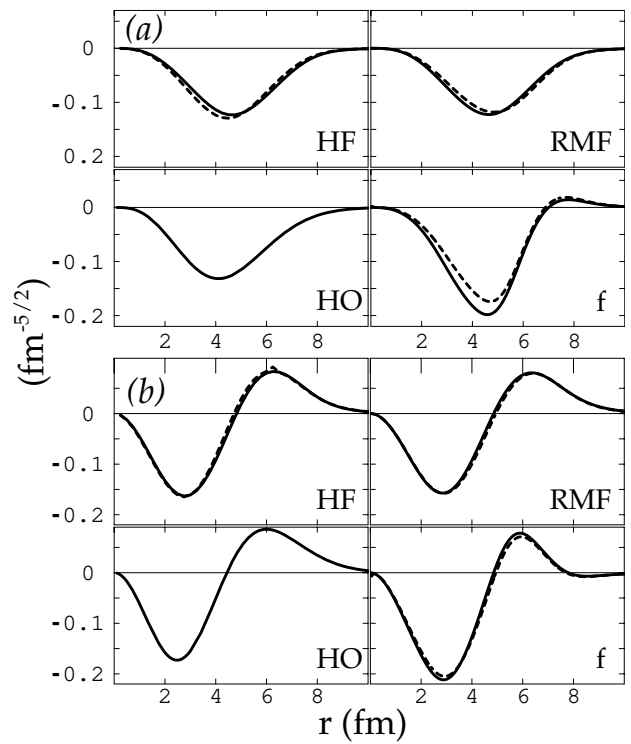}
\end{center}
\caption{$1f$ (a) and $2d$ (b) spin partners' wave functions of $^{208}$Pb obtained in
HF,   HO, and
RMF  calculations. The plot labeled `f' shows the scaled values 
 of $D_{n\ell j=\ell-{1 \over 2}}g_{n\ell j=\ell-{1 \over 2}}$ (dashed line) and $D_{n\ell j=\ell+{1 \over 2}}g_{n\ell j=\ell+{1 \over 2}}$ (solid
line). See text for details.
}
\label{comparison1s:pic}
\end{figure}

In Figs.~\ref{comparison1s:pic} (a) and (b) we plot
the upper components $g_{n\ell j = \ell + {1 \over 2}}(r)$ (dashed line) and $g_{n\ell j = \ell - {1 \over 2}}(r)$ (solid line) using
the HO model, the non-relativistic HF approximation, and the
RMF approximation. The Nilsson model (HO) shows perfect agreement of course since it has a
constant spin-orbit potential. However even the self-consistent
non-relativistic and relativistic mean fields show very little
difference between eigenstates of the spin doublets.

In the lower right-hand part of Figs.~\ref{comparison1s:pic} (a) and (b), we compare  $\lambda_{n\ell
j={\ell}-{1\over2}} D_{{ n}{\ell} j= { \ell} - {1\over 2}}(r)\
f_{{ n}{\ell}j={\ell}-{1\over2}}(r)$ with $\lambda_{n\ell
j={\ell}+{1\over2}}D_{n\ell j= { \ell} + {1\over 2}}(r)f_{{ n}{
\ell} j={\ell}+{1\over2}}(r)$ where the factor of $\lambda_{n\ell
j}=({V_S(0) + V_V(0)+ E_{n\ell j}})^{-1}$ scales the expression to
be comparable in magnitude to the upper components, according to
equation (\ref{Dirac2}). The agreement for these differential
relations is also very good.

\section{Summary and Conclusions}
\label{sec5}

In the pseudospin symmetry limit the radial wave functions of the upper
components of pseudospin doublets satisfy certain differential
relations. We demonstrated that these relations are not only approximately
valid for the relativistic mean
field eigenfunctions but also for the non-relativistic Hartree-Fock and harmonic oscillator 
eigenfunctions. Generally, we expect
them to be approximately valid for eigenfunctions of any
non-relativistic phenomenological nuclear
potential that fits the spin-orbit splittings of nuclei.
Likewise in the spin symmetry limit the
radial amplitudes of the upper components of the Dirac eigenfunctions
of spin doublets are
predicted to be equal and this is approximately valid for both
non-relativistic and
relativistic mean field models. Also the spatial amplitudes of the
lower components of the
Dirac eigenfunctions of spin doublets satisfy differential
relations in spin symmetry limit and these relations are
approximately valid in the relativistic mean field model.

Hence we seem to have both spin and pseudospin dynamic symmetry; that
is, the energy
levels are not degenerate but the eigenfunctions well preserve both
symmetries. For both of these
symmetries to be conserved both the vector and scalar potentials must
be constant. Of
course this is not true. However, for heavy nuclei this is
approximately true in the
nuclear interior and exterior. Only on the surface are the potentials
changing rapidly.
This leads to a dynamic symmetry for
both spin and pseudospin. The spin-orbit splittings  are determined
by $d[V_V(r) - V_S(r)]/dr$
while the
pseudospin-orbit splittings are determined by $d[V_V(r) + V_S(r)]/dr$. Therefore, the energy
splittings for spin doublets
are larger than  for pseudospin doublets
because $V_V(r) - V_S(r)$ changes more rapidly on the nuclear surface than
$V_S(r) + V_V(r)$ because  $V_V(r) - V_S(r) \gg  |V_S(r) + V_V(r)|$
in the interior and both go to zero
in the nuclear exterior. 

\begin{acknowledgments}
Useful suggestions from Peter von Bretano are gratefully acknowledged. This work was supported in part by the 
U.S.~Department of Energy under Contract Nos.~DE-FG02-96ER40963 (University of Tennessee), 
DE-AC05-00OR22725 with 
UT-Battelle, LLC (Oak Ridge National Laboratory), ~W-7405-ENG-36 (Los Alamos), and by the Polish Comittee for Scientific Research (KBN) under Contract No.~5~P03B~014~21.
\end{acknowledgments}
\strut

\strut


\end{document}